\documentclass[11pt]{article}
\usepackage{amsmath}
\usepackage{amsfonts}
\usepackage{amssymb}
\usepackage{graphicx}
\usepackage{bm}
\usepackage{color}
\usepackage{amscd}

\def\bea{\begin{eqnarray}}
\def\eea{\end{eqnarray}}

\begin{document}
\begin{center}
\LARGE {\bf Black hole entropy in the Chern-Simons-like theories of gravity and Lorentz-diffeomorphism Noether charge }
\end{center}

\begin{center}
{M. R. Setare \footnote{E-mail: rezakord@ipm.ir}\hspace{1mm} ,
H. Adami \footnote{E-mail: hamed.adami@yahoo.com}\hspace{1.5mm} \\
{\small {\em  Department of Science, University of Kurdistan, Sanandaj, Iran.}}}\\

\end{center}

\begin{center}
{\bf{Abstract}}\\
In the first order formalism of gravity theories, there are some theories which are not Lorentz-diffeomorphism covariant. In the framework of such theories we cannot apply the method of conserved charge calculation used in Lorentz-diffeomorphism covariant theories. In this paper we firstly  introduce the total variation of a quantity due to an infinitesimal Lorentz-diffeomorphism transformation. Secondly, in order to obtain the conserved charges of Lorentz-diffeomorphism non-covariant theories, we extend the Tachikawa method \cite{3}. This extension includes not only Lorentz gauge transformation but also the diffeomorphism. We apply this method to the Chern-Simons-like theories of gravity (CSLTG) and obtain a general formula for the entropy of black holes in those theories. Finally, some examples on CSLTG are provided and the entropy of the BTZ black hole is calculated in the context of the examples.
\end{center}

\section{Introduction}
There is a class of gravitational theories in $(2+1)$-dimension (e.g. Topological massive gravity (TMG) \cite{1'}, New massive gravity (NMG)\cite{2'}, Minimal massive gravity (MMG) \cite{3'}, Zewi-dreibein gravity (ZDG)\cite{4'}, Generalized minimal massive gravity (GMMG) \cite{5'}, etc), called the Chern-Simons-like theories of gravity \cite{1}. In this work, we try to obtain a general expression for the entropy of the black hole solutions in the context of CSLTG. \\
In the metric formalism of gravity for the covariant theories defined by a Lagrangian $n$-form $L$, Wald has shown that the entropy of black holes is the Noether charge associated with the horizon-generating Killing vector field evaluated at the bifurcation surface\cite{2}. Presence of the purely gravitational Chern-Simons terms and mixed gauge gravitational ones gives rise to a non-covariant theory of gravity in the metric formalism. Tachikawa extended the Wald approach to include non-covariant theories. Hence, regarding this extension one can calculate the black hole entropy as a Noether charge in the context of non-covariant theories as well \cite{3}. Another way (apart of the Tachikawa method) to obtain the entropy of black holes in the context of such theories has been studied in the papers \cite{100}-\cite{500}, by an appropriate way. \\
 It is known that the Lagrangian of the CSLTG can be written in the first order formalism, in which the spin-connection is considered as an independent Lorentz vector valued one-form. Applying the Wald method to calculate the Noether charges in the first order formalism, one can find that the Noether charges are proportional to $\xi$, where $\xi$ is a Killing vector field corresponding to the conserved charge. It is clear that $\xi$ must be zero on the bifurcation surface when we calculate the entropy of black hole, because $\xi$ is the horizon-generating Killing vector field which is zero on the bifurcation surface. It seems disappointing at the first glance because it appears that the entropy will be zero, but it is not true. Recently, it has been shown that in approaching to the bifurcation surface, the spin-connection diverges in a way that the spin-connection interior product in $\xi$  remains finite ensuring that there is no problem \cite{4}. Another way to deal with this problem was proposed in \cite{6}, in which the authors chose the Cauchy surface where the event horizon does not lie on bifurcation surface. To avoid any confusion and for extending  the Wald approach to include Lorentz invariance in addition to diffeomorphism invariance, Jacobson and Mohd  have introduced the so-called Lorentz-Lie derivative \cite{4}. Lorentz-Lie derivative is a generalization of the Lie derivative, and it is  covariant under the Lorentz-diffeomorphism transformations. The authors of  \cite{4} demand that the Lorentz-Lie derivative vanishes when $\xi$ is a Killing vector filed. \\
It is clear that the CSLTG are manifestly diffeomorphism covariant theories but their Lagrangian may be non-invariant under the Lorentz gauge transformations. In this work we try to extend the Tachikawa method to be able to calculate Noether charges of a theory which is not covariant under the general Lorentz-diffeomorphism transformations. \\ We obtain a generic formula for the entropy of all stationary black hole solutions of any Chern-Simons-like theory of gravity, such as TMG, NMG, GMG, MMG, GMMG, ZDG, etc. It is interesting that our formula is very simple. One only need to know the coupling constants and the field content of the model, specially for the stationary black hole solution of such theories, where the horizon of black hole is a circle.\\
We use lower case Greek letters for the spacetime indices, and the internal Lorentz indices are denoted by lower case Latin letters. The metric signature is mostly plus.

\section{Lorentz-Lie derivative and total variation}
Suppose that the dimension of spacetime is $n$. Let $e^{a}_{\hspace{1.5 mm} \mu}$ denote the vielbein. Under a Lorentz gauge transformation, $e^{a}_{\hspace{1.5 mm} \mu}$  transforms as $\tilde{e}^{a}_{\hspace{1.5 mm} \mu} = \Lambda ^{a}_{\hspace{1.5 mm} b} e^{b}_{\hspace{1.5 mm} \mu}$ where $\Lambda \in SO(n-1,1) $, i.e. $ e^{a}= e^{a}_{\hspace{1.5 mm} \mu}dx^{\mu}$ is $SO(n-1,1) $ vector valued 1-form. The Lorentz-Lie derivative (L-L derivative) of $e^{a}$ defined as follow \cite{4}:
\begin{equation}\label{1}
  \mathfrak{L}_{\xi} e^{a} = \pounds_{\xi} e^{a} +\lambda ^{a}_{\hspace{1.5 mm} b} e^{b} ,
\end{equation}
where $\pounds_{\xi}$ denote ordinary Lie derivative along $\xi$ and $\lambda ^{a}_{\hspace{1.5 mm} b}$ generates the Lorentz gauge transformations $SO(n-1,1)$. In general, $\lambda ^{a}_{\hspace{1.5 mm} b}$ is independent from the vielbein and spin-connection and is a function of space-time coordinates and of the diffeomorphism generator $\xi$. It is straightforward extension of this expression of the L-L derivative for $e^{a}$ to the case for which we have more than one Lorentz index. If we demand that the L-L derivative of the Minkowski metric $\eta _{ab}$ vanishes then we find that $\lambda _{ab}$ must be antisymmetric. In order to the L-L derivative be covariant under the Lorentz transformations, $\lambda ^{a}_{\hspace{1.5 mm} b}$ must transforms like a connection for the L-L derivative, $\tilde{\lambda}= \Lambda \lambda \Lambda ^{-1}+ \Lambda \pounds_{\xi} \Lambda ^{-1}$. \\
We know that under Lorentz transformation the one-form spin-connection $\omega ^{a}_{\hspace{1.5 mm} b} = \omega ^{a}_{\hspace{1.5 mm} b \mu} dx^{\mu}$ transforms as $\tilde{\omega}= \Lambda \omega \Lambda ^{-1}+ \Lambda d \Lambda ^{-1}$, then, although it is an invariant quantity under diffeomorphism but it is not invariant under Lorentz gauge transformation. \\
Now we introduce the total variation as combination of variations due to the diffeomorphism and the infinitesimal Lorentz transformation, i.e. $\delta _{\xi}=\delta _{diffeo}+\delta _{Lorentz}$. It is obvious that the total variation of $e^{a}$ is equal to its L-L derivative, $\delta _{\xi} e^{a} = \mathfrak{L}_{\xi} e^{a}$. One can calculate the total variation of the spin-connection and since it is not a Lorentz-diffeomorphism invariant quantity then we obtain
\begin{equation}\label{2}
  \delta _{\xi} \omega ^{ab} = \mathfrak{L}_{\xi} \omega ^{ab} -d \lambda ^{ab}.
\end{equation}
To avoid the appearance of extra term $ -d \lambda ^{ab} $, in the above equation, the authors in \cite{4} have defined  the L-L derivative of spin-connection exceptionally. But it is interesting that here we have a general definition, without exception, for the L-L derivative, so the equation \eqref{2} emphasizes that the spin-connection is not a Lorentz-diffeomorphism invariant quantity. Therefore, it is clear that the CSLTG can be a non-covariant theory under a general Lorentz-diffeomorphism transformation and this assure that we must generalize the Tachikawa approach to include the Lorentz gauge transformations in addition to diffeomorphism.

\section{Extended Tachikawa method}
Now consider the Lagrangian $n$-form $L(\Phi)$, where $\Phi$ is collection of dynamical fields. The arbitrary variation of this lagrangian is given by
\begin{equation}\label{3}
  \delta L = E_{\Phi} \delta \Phi + d \Theta (\Phi, \delta \Phi),
\end{equation}
where $E_{\Phi}=0$ are equations of motion of the theory and $ \Theta (\Phi, \delta \Phi) $ is the symplectic potential $(n-1)$-form which is linear in $\delta \Phi$. We suppose that this Lagrangian is not invariant under Lorentz-diffeomorohism transformations, so its total variation will have the following form
\begin{equation}\label{4}
  \delta _{\xi} L = \mathfrak{L}_{\xi} L +d \psi _{\xi} ,
\end{equation}
where $ \psi _{\xi}$ is a suitable $(n-1)$-form. Since $\mathfrak{L}_{\xi} L = \pounds _{\xi} L = d i_{\xi} L $, where $ i_{\xi} $ denotes exterior derivative in $\xi$, we can define the current $(n-1)$-form as follow
\begin{equation}\label{5}
  j _{\xi} = \Theta (\Phi, \delta_{\xi} \Phi) - i _{\xi} L - \psi _{\xi} ,
\end{equation}
for which we have $ d j _{\xi} = - E_{\Phi} \delta \Phi $. Therefore, $j_{\xi}$ is indeed a conserved current on-shell, i.e. $d j _{\xi} \simeq 0 $, where $\simeq$ emphasizes that the equality holds just on-shell. So, $j_{\xi}$ is a closed form on-shell and \cite{5} implies that it is an exact form on-shell, so one can write $j_{\xi} \simeq d Q _{\xi}$, where $ Q _{\xi} $ is a conserved charge $(n-2)$-form on-shell. The total variation of $\Theta (\Phi, \delta \Phi) $ can be written as
\begin{equation}\label{6}
  \delta _{\xi} \Theta (\Phi, \delta \Phi) = \mathfrak{L}_{\xi} \Theta (\Phi, \delta \Phi) + \Pi _{\xi} ,
\end{equation}
 then by calculating $\delta \delta _{\xi} L$ in two ways, we obtain $ d \Pi _{\xi} \simeq \delta  d \psi _{\xi} $. Thus by virtue of \cite{5}, we can write
\begin{equation}\label{7}
   \Pi _{\xi} - \delta \psi _{\xi} \simeq d \Sigma _{\xi} .
\end{equation}
If we take an arbitrary variation in \eqref{5}, by some calculations we can show that
\begin{equation}\label{8}
  \begin{split}
     \Omega (\Phi, \delta \Phi , \delta _{\xi} \Phi ) & = \delta \Theta (\Phi, \delta _{\xi} \Phi) - \delta  _{\xi} \Theta (\Phi, \delta \Phi) \\
       & \simeq \delta j_{\xi} -d i_{\xi} \Theta (\Phi, \delta \Phi)- d \Sigma _{\xi} ,
  \end{split}
\end{equation}
 where the first equality is definition of $(n-1)$-form symplectic current $\Omega$  which is an anti-symmetrized field variation of $\Theta $ and it is linear in $\delta \Phi$ and $\delta _{\xi} \Phi$. Let us define $Q^{\prime}_{\xi} $ as follow \footnote{In definition of $Q ^{\prime}_{\xi}$ we assume that one can write down $i_{\xi} \Theta (\Phi, \delta \Phi)+ \Sigma _{\xi}$ as a variation of a (n-2)-form, say $C_{\xi}$, that is $i_{\xi} \Theta (\Phi, \delta \Phi)+ \Sigma _{\xi} = \delta C_{\xi} $.  }
\begin{equation}\label{9}
  \delta Q ^{\prime}_{\xi} = \delta Q _{\xi} - i_{\xi} \Theta (\Phi, \delta \Phi)- \Sigma _{\xi} ,
\end{equation}
then using this definition and $j_{\xi} \simeq d Q _{\xi}$, one can rewrite the equation \eqref{8} in the following form
\begin{equation}\label{10}
   \Omega (\phi, \delta \Phi , \delta _{\xi} \Phi ) \simeq \delta d Q ^{\prime}_{\xi} .
\end{equation}
It should be noted that here dependence of $\lambda_{ab}$ on $\xi$ can be determined so that $Q ^{\prime}_{\xi}$ still remains linear in $\xi$. On the other hand, the variation of Hamiltonian associated with $\xi$ defined as $\delta H_{\xi}= \int_{\mathcal{C}}  \Omega (\phi, \delta \phi , \delta _{\xi} \phi ) $, where $\mathcal{C}$ is a Cauchy surface. Hence, assuming that $t$  and $\phi$ are generators of the global time translation and the angular rotation, respectively, then the "canonical energy" $\mathcal{E}$ and the "canonical angular momentum" $\mathcal{J}$ are as following
\begin{equation}\label{11}
   \mathcal{E} = \int_{\infty} Q ^{\prime}_{t} , \hspace{5 mm}  \mathcal{J} = - \int_{\infty} Q ^{\prime}_{\phi} ,
\end{equation}
respectively, where the integrals are taken over an $(n-2)$-dimensional sphere at infinity. \\
 If we have a stationary black hole spacetime with a bifurcate Killing horizon generated by $\xi = t + \Omega _{H} \phi $, where $\Omega _{H} $ is the angular velocity of the horizon, then we will have $\delta _{\xi} \Phi =0 $. Since the symplectic current $\Omega$ is linear in $\delta _{\xi} \Phi $, therefore by integrating over $\Omega (\phi, \delta \Phi , \delta _{\xi} \Phi )$, in Eq. \eqref{10} on a Cauchy surface, one gets $\int_{\partial \mathcal{C}} Q^{\prime}_{\xi} \simeq 0$, since $Q^{\prime}_{\xi}$ is linear in $\xi$. Then using \eqref{11} one can rewrite this as follow
\begin{equation}\label{12}
  \delta \int_{\mathcal{B}} Q^{\prime}_{\xi} = \delta \mathcal{E} - \Omega _{H} \delta \mathcal{J} ,
\end{equation}
where $\mathcal{B}$ is the bifurcation surface. Since $ Q^{\prime}_{\xi} $ is linear in $\xi$ and $\nabla \xi$ (due to $ \lambda_{ab} $), on the bifurcation surface we have \cite{4}
\begin{equation}\label{13}
  \xi ^{\mu}=0 , \hspace{7 mm} \nabla ^{\mu} \xi ^{\nu} = \kappa n^{\mu \nu} ,
\end{equation}
where $ \kappa $ and $ n^{\mu \nu} $ are the surface gravity and bi-normal to $\mathcal{B}$, respectively. Now we can define black hole entropy as
\begin{equation}\label{14}
  S = 2 \pi \int_{\mathcal{B}} \hat{Q}^{\prime}_{\xi} ,
\end{equation}
where $\hat{Q}^{\prime}_{\xi}=Q^{\prime}_{\xi} \vert _{\xi \rightarrow 0, \nabla \xi \rightarrow n} $. So the equation \eqref{12} represent the first law of black hole mechanics $T_{H} \delta S = \delta \mathcal{E} - \Omega _{H} \delta \mathcal{J}$, where $T_{H}= \kappa /(2 \pi)$.

\section{Black hole entropy in Chern-Simons-like theories of gravity}
It is known that a Chern-Simons-like theory of gravity is a theory in $(2+1)$-dimension and its Lagrangian 3-form is given by \cite{1}
 \begin{equation}\label{15}
  L=\frac{1}{2} g_{rs} u^{r} \cdot d u^{s}+\frac{1}{6} f_{rst} u^{r} \cdot u^{s} \times u^{t},
\end{equation}
where $ u^{ra}=u^{ra}_{\hspace{3 mm} \mu} dx^{\mu} $ are Lorentz vector valued one-forms, $r$ ($r=1,...,N$) and $a$ refer to flavour and Lorentz indices, respectively. Also, $g_{rs}$ is a symmetric constant metric on the flavour space and $f_{rst}$ is a totally symmetric "flavour tensor" which is interpreted as the coupling constants. \\
We know that $ u^{ra} =\{ e^{a}, \omega ^{a} , h^{a} , \cdots \} $, where $e^{a}$, $\omega ^{a}= \frac{1}{2} \varepsilon ^{abc} \omega _{bc}$ and $ h^{a}_{\hspace{1.5 mm} \mu} = e^{a}_{\hspace{1.5 mm} \nu} h^{\nu}_{\hspace{1.5 mm} \mu} $ are dreibein, dualized spin-connection and auxiliary field, respectively. One can easily show that $\mathfrak{L}_{\xi} \varepsilon ^{abc}=0$, so using \eqref{2} we have $\delta _{\xi} \omega ^{a} = \mathfrak{L}_{\xi} \omega ^{a} -d \chi ^{a}$, where $\chi ^{a}= \frac{1}{2} \varepsilon ^{abc} \lambda _{bc} $. Therefore, we can write down the total variation of $ u^{rb} $ as follow
\begin{equation}\label{16}
  \delta _{\xi} u^{ra} = \mathfrak{L}_{\xi} u^{ra} -\delta ^{r} _{\omega} d \chi ^{a} ,
\end{equation}
where $ \delta ^{r} _{s} $ is Kronecker delta. Thus, it is obvious that apart from the diffeomorphism covariance, these theories may be non-covariant under the Lorentz gauge transformations and then for calculating black hole entropy, we must use of the method which described in the previous section. The variation of the Lagrangian \eqref{15} is
\begin{equation}\label{17}
  \delta L = \delta u^{r} \cdot E_{r} + d \Theta (u, \delta u),
\end{equation}
where
\begin{equation}\label{18}
   E_{r}^{\hspace{1.5 mm} a} = g_{rs} d u^{sa} + \frac{1}{2} f_{rst} (u^{s} \times u^{t})^{a} ,
\end{equation}
\begin{equation}\label{19}
   \Theta (u, \delta u) = \frac{1}{2} g_{rs} \delta u^{r} \cdot u^{s} .
\end{equation}
On the one hand, for all of our interesting Chern-Simons-like theories of gravity we have $ f_{\omega rs} = g_{rs}$ \cite{7}, also unlike the ordinary Lorentz derivative, the L-L derivative do not commute with exterior derivative, then by calculating \eqref{4} for this case, we can find that
\begin{equation}\label{20}
  \psi _{\xi} = \frac{1}{2} g_{\omega r} d \chi \cdot u^{r} .
\end{equation}
By taking total variation of $\Theta (u, \delta u)$ and comparing the result with \eqref{6} we obtain
\begin{equation}\label{21}
   \Pi _{\xi} = \frac{1}{2} g_{\omega r} d \chi \cdot \delta u^{r} .
\end{equation}
By substituting \eqref{20} and \eqref{21} into \eqref{7} one gets $d \Sigma _{\xi} \simeq 0$ and then we can choose $\Sigma _{\xi} \simeq 0$.
Since the current is defined by \eqref{5} we can now obtain that for the CSLTG:
\begin{equation}\label{22}
   j _{\xi} = d Q_{\xi} - i_{\xi} u^{r} \cdot E_{r} + \chi \cdot E_{\omega} ,
\end{equation}
where
\begin{equation}\label{23}
  Q_{\xi} = \frac{1}{2} g_{rs} i_{\xi} u^{r} \cdot u^{s} - g_{\omega r} \chi \cdot u^{r} ,
\end{equation}
as we expect $j _{\xi}$ is an exact form on-shell. By substituting \eqref{19} and \eqref{23} into \eqref{9} we have
\begin{equation}\label{24}
 \delta Q_{\xi}^{\prime} = ( g_{rs} i_{\xi} u^{s} - g_{\omega r} \chi ) \cdot \delta u^{r}.
\end{equation}
Then by integrating over the above formula on the bifurcation surface one gets
\begin{equation}\label{25}
 \int_{\mathcal{B}} Q_{\xi}^{\prime} = - g_{\omega r} \int_{\mathcal{B}} \chi  \cdot u^{r} .
\end{equation}
So far, we take $\lambda ^{ab}$ as a function of space-time coordinates and of the diffeomorphism generator $\xi$ and it is antisymmetric with respect to $a$ and $b$. To obtain an explicit expression for $\lambda ^{ab}$, in an appropriate manner, the authors in \cite{4} demand that it must be chosen so that
the L-L derivative of $e^{a}$ vanishes when $\xi$ is a Killing vector field and then they showed that $\lambda ^{ab}$ should be provided as follows:
\begin{equation}\label{26}
 \lambda ^{ab} = e^{\sigma [a} \pounds _{\xi} e^{b]}_{\hspace{1.5 mm} \sigma}  .
\end{equation}
Since $\chi ^{a}= \frac{1}{2} \varepsilon ^{abc} \lambda _{bc} $, then using above formula one can show that
\begin{equation}\label{27}
  \chi ^{a} = i_{\xi} \omega ^{a} -\frac{1}{2} \varepsilon ^{a}_{\hspace{1.5 mm} bc} e^{\nu b} (i_{\xi} T^{c})_{\nu} + \frac{1}{2} \varepsilon ^{a}_{\hspace{1.5 mm} bc} e^{b}_{\hspace{1.5 mm} \mu} e^{c}_{\hspace{1.5 mm} \nu} \nabla ^{\mu} \xi ^{\nu} ,
\end{equation}
where $T^{a}$ is the torsion 2-form. Thus, on the bifurcation surface we will have $\chi^{a} \vert _{\mathcal{B}} =\kappa N^{a} $, where $N^{a}= \frac{1}{2} \varepsilon ^{abc} n _{bc}$. Therefore, using \eqref{14} and \eqref{25}, the black hole entropy in the Chern-Simons-Like theories of gravity is given by following formula:
\begin{equation}\label{28}
  S=- 2 \pi g_{\omega r} \int_{\mathcal{B}} N \cdot u^{r} .
\end{equation}
 We should mention that $N^{a}$ is normalized to $+1$ because $n^{\mu \nu}$ is normalized to $-2$. By specifying up $g_{rs}$ and $f_{rst}$, one can solve the equations of motion \eqref{18} for determining explicit form of $u^{r}$'s and then we can substitute $g_{rs}$ and $u^{r}$'s in the equation \eqref{28} for a given black hole spacetime solution, so we will have the entropy of black hole for the given special solution.

\section{The entropy of stationary black holes}

In the above section, we found a general formula for entropy of black holes in all of Chern-Simons-like theories of gravity. Consider a stationary black hole solution of such theories so that the horizon of black hole is a circle which is located at $r=r_{h}$ then, the non-zero components of bi-normal to this horizon are $n_{01}=-n_{10}$. Because $N^{a}=\frac{1}{2} \varepsilon ^{a b c} n_{b c}$ and $N^{a}$ is normalized to $+1$ then, the only non-zero component of $N^{a}$ is $N^{\phi}=(g_{\phi \phi})^{-1/2}$. Thus, the formula \eqref{28} takes the following form

\begin{equation}\label{29}
  S = - 2 \pi g_{\omega s} \int_{r=r_{h}} (g_{\phi \phi})^{-1/2} u^{s} _{\hspace{1.5 mm} \phi \phi } d \phi .
\end{equation}
Hence, this formula is applicable to calculate the entropy of the stationary black holes.

\section{Examples}
In this section, we calculate the entropy of BTZ black hole in the context of some models which are known as Chern-Simons-like theories of gravity.

\subsection{Minimal massive gravity}
Now, we use the formula \eqref{29} to calculate the entropy of BTZ black hole in the context of minimal massive gravity \cite{3'}. In this model, $u^{s}= \{ e,\omega , h \}$ where $e$, $\omega$ and $h$ are dreibein, dualized spin-connection and an auxiliary field, respectively. Also, the non-zero components of the flavour metric are $g_{e \omega}=-\sigma $, $g_{e h}=1 $ and $g_{\omega \omega}=1 / \mu $. As we know, this model is not a torsion-free theory, so we can write
\begin{equation}\label{30}
  \omega ^{a} _{\hspace{1.5 mm} \mu}=\Omega (e) ^{a} _{\hspace{1.5 mm} \mu} - \alpha h ^{a} _{\hspace{1.5 mm} \mu},
\end{equation}
so that
\begin{equation}\label{31}
\begin{split}
    & \Omega (e) ^{\mu} _{\hspace{1.5 mm} \nu} = \frac{1}{2} \epsilon ^{\mu \alpha \beta} e ^{c} _{\hspace{1.5 mm} \beta} \nabla _{\nu} e_{c \alpha} , \\
     & h _{\mu \nu} = - \frac{1}{\mu ( 1 + \alpha \sigma)^{2} } \left( S _{\mu \nu} + \frac{\alpha \Lambda _{0}}{2} g _{\mu \nu} \right),
\end{split}
\end{equation}
where $\Omega (e)$ is the usual torsion-free dualized spin-connection, $\epsilon _{\mu \alpha \beta}= \sqrt{-g} \varepsilon _{\mu \alpha \beta} $ is the Levi-Civita tensor and $S _{\mu \nu}=R _{\mu \nu}- \frac{1}{4} g _{\mu \nu} R $ is the 3D Schouten tensor. Hence, for this model, the formula \eqref{29} simply rewritten as
\begin{equation}\label{32}
  S = - 2 \pi \int_{r=r_{h}} (g _{\phi \phi})^{-\frac{1}{2}} \left( - \sigma g _{\phi \phi} + \frac{1}{\mu} \Omega (e) _{\phi \phi} -\frac{\alpha}{\mu} h _{\phi \phi} \right) d \phi .
\end{equation}
One can use this formula to calculate the entropy of the stationary black hole solutions of this model. Here, we apply this formula to obtain the entropy of the BTZ black hole. For the BTZ black hole solution, we have
\begin{equation}\label{33}
  \begin{split}
       & e^{0}=\left( \frac{(r^{2}-r_{+}^{2})(r^{2}-r_{-}^{2})}{l^{2}r^{2}} \right)^{\frac{1}{2}}dt \\
       & e^{1}= r \left( d \phi -\frac{r_{+}r_{-}}{lr^{2}} dt \right) \\
       & e^{2}=\left( \frac{l^{2}r^{2}}{(r^{2}-r_{+}^{2})(r^{2}-r_{-}^{2})} \right)^{\frac{1}{2}}dr
  \end{split}
\end{equation}
and we can find that
\begin{equation}\label{34}
  g _{\phi \phi} = r^{2} , \hspace{0.5 cm} \Omega (e) _{\phi \phi} = - \frac{r_{+} r_{-}}{l}, \hspace{0.5 cm} S_{\phi \phi} = -\frac{r^{2}}{2 l^{2}} , \hspace{0.5 cm} h _{\phi \phi} = \frac{(1 - \alpha \Lambda _{0} l^{2}) r^{2}}{2 \mu l^{2} (1+ \alpha \sigma )^{2} }.
\end{equation}

By substituting above results into \eqref{32}, we simply deduce the entropy of the BTZ black hole as follows:
\begin{equation}\label{35}
S = 4 \pi ^{2} \left( \sigma r_{+} + \frac{ r_{-}}{\mu l}
+ \frac{\alpha (1-\alpha \Lambda _{0} l^{2}) r_{+} }{2 \mu^{2} l^{2} (1+\alpha \sigma)^{2}} \right) ,
\end{equation}
which is exactly what has been found in \cite{6}. We see that the formula \eqref{28} or \eqref{29} is a general formula to calculate the entropy of black hole solutions of the CSLTG which present very simple form in comparsion with entropy formula by other methods.

\subsection{Generalized massive gravity}
In this sub-section, we consider generalized massive gravity as another example. This model first introduced in \cite{2'}, then studied more in \cite{16}. In this model, there are four flavours of one-form, $u^{s}= \{ e, \omega , h, f \}$ and the non-zero components of the flavour metric are $g_{e \omega}=-\sigma$, $g_{e h}=1$, $g_{\omega f}=-\frac{1}{m^{2}}$ and $g_{\omega \omega}=\frac{1}{\mu}$. This model is torsion-free, so the dualized spin-connection is given by $\omega ^{a} = \Omega ^{a} (e)$, also one can find that $f_{\mu \nu } = - S_{\mu \nu}$. Thus, for this model, one can easily rewrite \eqref{29} as follows:
\begin{equation}\label{36}
  S = - 2 \pi \int_{r=r_{h}} (g _{\phi \phi})^{-\frac{1}{2}} \left( - \sigma g _{\phi \phi} + \frac{1}{\mu} \Omega (e) _{\phi \phi} +\frac{1}{m^{2}} S _{\phi \phi} \right) d \phi  .
\end{equation}
By substituting \eqref{34} into the above formula,one can easily find that the entropy of BTZ black hole in generalized massive gravity is given by
\begin{equation}\label{37}
S = 4 \pi ^{2} \left[ \left( \sigma + \frac{1}{2 m^{2} l^{2}} \right) r_{+} + \frac{ r_{-}}{\mu l} \right] .
\end{equation}
 After a re-parametrization of parameters, this is what have been given in \cite{9}.

\subsection{New version of generalized zwei-dreibein gravity}

As the last example, we consider the new version of Generalized zwei-dreibein gravity (GZDG$^{+}$) which is recently proposed and investigated in \cite{8}. In the GZDG$^{+}$, there are five flavours of one-form, $u^{s}= \{ e_{1}, \omega _{1} , e_{2}, \omega _{2}, h \}$. One can identified $e_{1}$ with the physical one, i.e. $e_{1} \equiv e $ hence, $\omega _{1} = \omega$. Because this model is torsion-free then we have $\omega = \Omega (e)$. On the other hand, in this model, the non-zero components of the flavour metric are $g_{e_{1} \omega_{1}}=-\sigma$, $g_{e_{2} \omega_{2}}=1$, $g_{e_{1} h}=-\sigma$ and $g_{ \omega_{1} \omega_{1}}=-\sigma$. Then, the entropy formula \eqref{29} reduces to the following one
\begin{equation}\label{38}
  S = - 2 \pi \int_{r=r_{h}} (g _{\phi \phi})^{-\frac{1}{2}} \left( - \sigma g _{ \phi \phi} + \frac{1}{\mu} \Omega (e) _{\phi \phi} \right) d \phi  ,
\end{equation}
 and by substituting \eqref{34} into this expression, the entropy of the BTZ black hole in the GZDG$^{+}$ is
 \begin{equation}\label{39}
S = 4 \pi ^{2} \left( \sigma r_{+} + \frac{ r_{-}}{\mu l} \right).
\end{equation}
It is clear that this is exactly the BTZ black hole entropy in the context of usual topologically massive gravity \cite{9'} (see also\cite{10}). This result do not surprise us because the integrand in \eqref{38} depends on metric itself and usual dualized spin-connection, explicitly. Actually, this feature is valid to all stationary black hole solutions of this model, in other words, the entropy of stationary black hole solutions in the GZDG$^{+}$ is exactly equal with entropy appear in topologically massive gravity.

\section{Conclusion}
Main models of massive gravity in $(2+1)$-dimension, such as TMG, MMG, GMMG, etc, contain the Chern-Simons (CS) term explicitly. Due to the presence of CS term, the Lagrangian of these models are not covariant, so we cannot apply the Wald formula to obtain the entropy of black holes in the context of these types of models. On the other hand, the Wald approach works for covariant Lagrangians constructed from the metric, but all CSLTG have been written in first order formalism. Furthermore, there are some Chern-Simons-like theories of gravity are not covariant under Lorentz-diffeomorphism transformations. Here, we have provided a unified approach to obtain the entropy of black hole solutions of all CSLTG. We were able to obtain a generic and interesting formula for the entropy of black hole solutions of CSLTG by Eq.(\ref{28}) which depends only on the fields $u^r$ and coupling constants $g_{r\omega}$ of the theory.
  Firstly, we have considered the Lorentz-Lie derivative and then we have introduced the total variation as a combination of variations due to the diffeomorphism and the infinitesimal Lorentz transformation. We have shown that the Lorentz-Lie derivative of spin-connection is not covariant under Lorentz gauge transformation, and as a result, theory is non-covariant under Lorentz-diffeomorphism transformation. Then in section $3$, we have extended the Tachikawa method which is an approach for calculating the conserved charges of diffeomorphism non-covariant theories. This extension  includes Lorentz gauge transformation in addition to diffeomorphism. In section $4$, we have applied this method to calculate conserved charges of the Chern-Simons-like theories of gravity as an example of a Lorentz-diffeomorphism non-covariant theory. Then we have extracted a general formula for black hole entropy in those theories. In section $5$, we have considered the stationary black hole solutions and have reduced Eq.\eqref{28} to an elegant formula Eq.\eqref{29}, for the entropy of stationary black holes. Finally, in section $6$, we have employed  Eq. \eqref{29} to obtain the entropy of BTZ black hole in the context of some models which are known as Chern-Simons-like theories of gravity.

\section{Acknowledgments}
M. R. Setare  thanks Dr. A. Sorouri  for his help in improvement the English of the text.

\end{document}